\documentclass[aps,reprint,amsmath,amssymb,pra]{revtex4-1}
\usepackage{amsmath}
\usepackage{graphicx}
\usepackage{dcolumn}
\usepackage{bm}
\usepackage[english]{babel} 
\usepackage[T1]{fontenc}

\usepackage[usenames,dvipsnames]{xcolor}
\usepackage[colorlinks=true,citecolor=Cerulean,linkcolor=RubineRed,urlcolor=Cerulean]{hyperref}

\newcommand{\ket}[1]{|{#1}\rangle}

\begin{document}

\title{Staggered Ground States in an Optical Lattice}
\author{Dean Johnstone$^1$, Niclas Westerberg$^{1,2}$, Callum W. Duncan$^1$ and Patrik \"{O}hberg$^1$}
\affiliation{$^1$SUPA, Institute of Photonics and Quantum Sciences,
	Heriot-Watt University, Edinburgh, EH14 4AS, UK \\ $^2$School of Physics \& Astronomy, University of Glasgow, Glasgow G12 8QQ, UK}

\date{\today}

\begin{abstract}
Non-standard Bose-Hubbard models can exhibit rich ground state phase diagrams, even when considering the one-dimensional limit. Using a self-consistent Gutzwiller diagonalisation approach, we study the mean-field ground state properties of a long-range interacting atomic gas in a one-dimensional optical lattice. We first confirm that the inclusion of long-range two-body interactions to the standard Bose-Hubbard model introduces density wave and supersolid phases. However, the introduction of pair and density-dependent tunnelling can result in new phases with two-site periodic density, single-particle transport and two-body transport order parameters. These staggered phases are potentially a mean-field signature of the known novel twisted superfluids found via a DMRG approach [PRA \textbf{94}, 011603(R) (2016)]. We also observe other unconventional phases, which are characterised by sign staggered order parameters between adjacent lattice sites.
\end{abstract}

\maketitle

\section{Introduction}
The Bose-Hubbard model and its extensions has long been a theoretical workhorse for lattice based systems \cite{Gutzwiller1963,kanamori1963,hubbard1963,hubbard1964}, including in the field of ultracold atoms in optical lattices \cite{Jaksch1998,bloch2005,lewenstein2007,Dutta2015}. One of the first topics of great interest to the ultracold gas community was the Mott-insulator to superfluid transition in the standard Bose-Hubbard model \cite{Fisher1989,Freericks1994,greiner2002,Zwerger2003,bakr2010}. Ultracold gases in optical lattices allows for full control of the underlying periodicity and inherently produces no defects. In addition, several techniques also exist that can tune the strength of interaction and correlation processes such as Feshbach resonances \cite{Fedichev1996,inouye1998,Theis2004,PhysRevA.71.043604} and laser assisted tunnelling \cite{Miyake2013,PhysRevA.98.053628}. The standard Bose-Hubbard model is known to be a poor approximation for systems with strong, non-trivial long-range interactions, as can be realised in dipolar atomic species \cite{doi:10.1063/1.2839130, PhysRevA.97.060701,Griesmaier_2007}. Such additions to the standard Bose-Hubbard model are referred to as extended, or non-standard, Bose-Hubbard models, where long-range phenomena can significantly change the properties of the system \cite{PhysRevA.96.063611}. An example of a long-range interacting atomic gas is the case of dipolar atoms \cite{Aikawa2012,PhysRevA.85.033607,Lu2012}, for which a long-range dipole-dipole interaction is present that decays as $1/r^3$ \cite{Trefzger2011}.

 For long-range interactions, non-standard terms can include a density-dependent tunnnelling, pair tunnelling, and/or inter-site interactions. It is known that introducing a density-dependent tunnelling changes the critical point of the Mott-insulator to superfluid transition \cite{Maik_2013}. In addition to changing ground state properties, this term also affects the dynamics of the system \cite{Lkacki2013a}. For dipolar interactions the additional non-standard pair tunnelling can even destroy the Mott-insulating domains and introduce new phases \cite{Dutta2012,Maik_2013}, including the pair superfluid. It is also known that the introduction of nearest-neighbour two-body interactions can induce density wave and supersolid ground states, which spontaneously break the translational symmetry of the lattice \cite{PhysRevLett.108.115301,Mosadeq2015,Rossini_2012,PhysRevB.95.195101,PhysRevB.86.054520}.

In this work, we will consider the ground state phases of atoms with long-range interactions in one-dimensional optical lattices in detail by a Gutzwiller mean-field approach. There are a significant number of works in the current literature which consider the constrained density-dependent or extended Bose-Hubbard model \cite{Rossini_2012,PhysRevB.95.195101,PhysRevB.86.054520,Trefzger2011,Biedron2018}. By including all terms, the resulting phase diagrams can differ significantly even for modest parameter strengths. We will confirm the destruction of the Mott-insulating phase and the introduction of known supersolid, density wave and pair superfluid phases \cite{srep12912}. By performing a detailed study of the ground state phase diagrams for various parameter regions, we find new staggered superfluid and supersolid phases, including sign staggered behaviour of the ordinary and pair superfluid and supersolid.

We will begin by defining the Bose-Hubbard model for a long-range interacting atomic gas, and discuss the derivation of the parameter strengths in terms of the well-known Wannier functions. In Sec.~\ref{_sc3}, we will discuss the mean-field approach used in this work. The various phases encountered in the course of this work are defined and discussed in Sec.~\ref{sec:Character}. We will then consider the phase diagrams for each non-standard Bose-Hubbard term being non-zero and the physically relevant case of combinations of non-zero additional terms in Sec.~\ref{_sc4}.

\section{Bose-Hubbard model}

In this section we will define the one-dimensional Bose-Hubbard model for atoms with long-range interactions. The many-body Hamiltonian in second quantised form of an ultracold gas in an optical lattice described by $V_{ext}(r)$ is given by
\begin{equation} \label{eq_fieldH}
\begin{aligned}
\mathcal{\hat{H}} = \int dr \hat{\Psi}^\dagger (r) \bigg( -\dfrac{\hbar^2}{2m}\nabla^2 + V_{ext}(r) - \mu \bigg) \hat{\Psi} (r) \\ + \, \dfrac{1}{2} \int dr dr' \hat{\Psi}^\dagger (r) \hat{\Psi}^\dagger (r') V_{int}(r,r') \hat{\Psi} (r) \hat{\Psi} (r'),
\end{aligned}
\end{equation}
where $\mu$ is the chemical potential, $ \hat{\Psi} (r) \, (\hat{\Psi}^\dagger (r)) $ are the bosonic annihilation (creation) field operators obeying the standard canonical commutation relations, $V_{int}(r,r')$ is the two-body interaction potential, and $m$ is the mass. 

We consider the case of long-range interacting atoms (e.g. dipolar atoms), which have an interaction potential of the form
\begin{equation}
V_{int}(r) = V_{S}(r) + V_{L}(r),
\end{equation}
with a short-range interaction, usually of contact-type, with
\begin{equation}
V_{S}(r) = g\delta(r),
\end{equation}
and a long-range interaction of the form
\begin{equation}
V_{L}(r) = \gamma h(r),
\end{equation}
where $g$ and $\gamma$ are scaling pre-factors and $h(r)$ is the non-local spatial profile of the interaction. As an example, for dipolar atoms, the non-local spatial profile is of form $h(r)=1/|r|^3$.

For a periodic external potential (e.g. an optical lattice), the continuous field operators may be described by discrete mode excitations by virtue of Bloch's theorem,
\begin{equation}
\hat{\Psi} (r) = \sum\limits_{n,k}\psi_{n,k}(r)\hat{b}_{n,k},
\label{eq:DiscreteMode}
\end{equation}
where $n$ labels the band, $k$ is the quasi-momentum, $\hat{b}_{n,k} \, (\hat{b}_{n,k}^\dagger)$ are the bosonic particle destruction (creation) operators, and $\psi_{n,k}(r)$ characterises the wave function. In the tight-binding limit, the delocalised $\psi_{n,k}(r)$ can be expressed in terms of localised and orthogonal Wannier functions, i.e.
\begin{equation}
\psi_{n,k}(r) = \sum\limits_{R}w_{n,R}(r)e^{i\textbf{k}.\textbf{R}},
\end{equation}
with $R$ corresponding to the lattice translation vector and $w_{n,R} (r)$ being the Wannier function in the $n$th band. Substituting the form in terms of Wannier functions into Eq.~\eqref{eq:DiscreteMode} gives
\begin{equation}
\hat{\Psi} (r) = \sum\limits_{n,R}w_{n,R}(r)\hat{b}_{n,R}.
\end{equation}
The overlap and extent of the Wannier functions is determined by the external potential's depth and is independent of the interaction strength and mechanism.

The usual tight-binding limit considers the lattice potential to be sufficiently deep such that the atoms are well localised to each lattice site, in an analogous way to electrons being tightly bound to atoms in solid state crystals \cite{Rosenberg1988}. In this limit, we can expand the wave function into the lowest set of Wannier functions. As the interaction and tunnelling terms usually decay substantially as the distance between sites increases, it is usually a good approximation to constrain the terms of the Hamiltonian to on-site and nearest-neighbour only. Using the Wannier representation, the general Bose-Hubbard model for long-range interacting atomic gases can be derived from Eq.~\eqref{eq_fieldH} as,
\begin{equation} \label{eq_ebh}
\begin{aligned}
\mathcal{\hat{H}} = & - J\sum\limits_{\langle i,j \rangle} \hat{b}_{i}^\dagger \hat{b}_{j} + \dfrac{U}{2} \sum\limits_{i} \hat{n}_i(\hat{n}_i - 1) \\ & - \mu\sum\limits_{i}\hat{n}_i  + \, V\sum\limits_{\langle i,j \rangle}\hat{n}_i\hat{n}_j  + \, P\sum\limits_{\langle i,j \rangle}\hat{b}_{i}^{\dagger 2} \hat{b}_{j}^2 \\ & + T\sum\limits_{\langle i,j \rangle} \hat{b}_{i}^\dagger (\hat{n}_i + \hat{n}_j ) \hat{b}_{j},
\end{aligned}
\end{equation}
where $i(j)$ are labels of the lattice sites, $\hat{n}_i = \hat{b}_{i}^\dagger \hat{b}_{i}$ is the number operator, and $\langle i,j \rangle$ indicates nearest-neighbour summations.

\begin{figure*}[t]
	\centering
	\includegraphics[width=\linewidth]{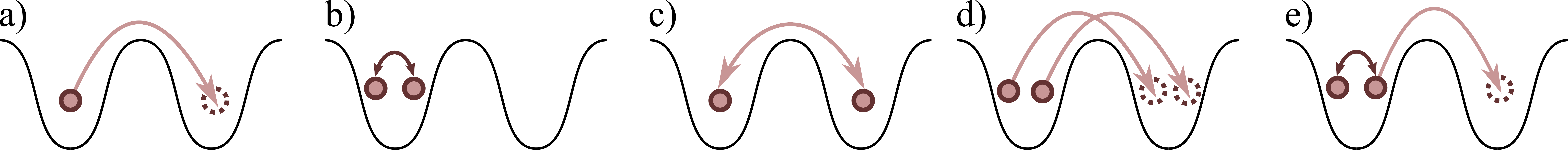}
	\caption{Illustrations of the two-body two-site terms contained in the density-dependent Bose-Hubbard Hamiltonian of Eq.~\eqref{eq_ebh}. a) The single-atom tunnelling $J$, b) the on-site two-body interaction $U$, c) the long-range two-site two-body interaction $V$, d) the pair tunnelling $P$, and e) the two-body density-dependent tunnelling $T$.}
	\label{fig:Illustrations}
\end{figure*}

In Hamiltonian~\eqref{eq_ebh} there are three terms which are contained in the standard Bose-Hubbard model; tunnelling with strength $J$, two-body on-site interactions of strength $U$, and a chemical potential of $\mu$. However, there are more exotic terms which describe other possible two-body processes. The first of these terms is the two-body nearest-neighbour interaction of strength $V$, which is vital when considering dipolar BECs. There is also a term which denotes the pair tunnelling process, i.e. when two atoms tunnel to a site together, which is of strength $P$. The final term is the density-dependent tunnelling, which is a single-particle tunnelling process that depends on the density of atoms at each site involved in the tunnelling process. We will denote the strength of the density-dependent term as $T$. The five dynamic tunnelling and interaction terms in Hamiltonian~\eqref{eq_ebh} are illustrated in Fig.~\ref{fig:Illustrations}.

Each of the coefficients for the terms of Hamiltonian~\eqref{eq_ebh} are defined in terms of overlap integrals of the Wannier functions. The single-particle tunnelling strength is given by
\begin{equation}
J = \dfrac{\hbar^2}{2m} \int dr  w_n^*(r) \nabla^2 w_{m}(r) - \int dr  w_n^*(r) w_{m}(r) V_{ext}(r),
\end{equation}
and the two-body on-site interaction by
\begin{equation}
U = g W_{nnnn}^S + \gamma W_{nnnn}^L,
\end{equation}
where
\begin{equation}
W_{ijkl}^S  = \int dr w_i^*(r) w_j^*(r) w_k(r) w_l(r),
\end{equation}
stands for the short-range interaction integral and 
\begin{equation}
W_{ijkl}^L  = \int dr dr' w_i^*(r) w_j^*(r') h(r-r') w_k(r) w_l(r'),
\end{equation}
for the long-range interaction integral and $ijkl$ are general labellings, with each being either $n$ or $m$ in the overlap integrals. We also get here the form of all three non-standard Bose-Hubbard terms, with the long-range two-body interaction given by
\begin{equation}
V = \dfrac{\gamma}{2} \big( W_{mnmn}^L+ W_{nmmn}^L \big),
\end{equation}
the density-dependent single-particle tunnelling by
\begin{equation}
T =  \gamma W_{mnnn}^L,
\end{equation}
and the two-body pair tunnelling by
\begin{equation}
P = \dfrac{\gamma}{2}  W_{mmnn}^L.
\end{equation}
We consider on-site and nearest-neighbour terms only, as all higher order terms will be small. This means we consider only the cases of $|m-n| = \{ 0,1\}$. All three non-standard Bose-Hubbard terms depend on the long-range interaction, and can, therefore, be important for the case of strong dipolar interactions. We note that the strength of these long-range two-body terms will generally follow the relation
\begin{align}
V > T > P.
\label{eq:Scaling}
\end{align}
Since $U$ depends on both $g$ and $\gamma$, the relative scaling of local interactions can be tuned with some freedom. All parameters in Hamiltonian~\eqref{eq_ebh} are dependent on combinations of the external and two-body interaction potentials, including an overlap integral of the Wannier functions. As an example, we will provide some approximate values for the dipolar Hubbard parameters. First, we consider an optical lattice potential of the form $V_{ext} = V_0 \sin^2(x/2)$, where $V_0$ is the lattice depth, and the units of energy and length are set to $E_R$ (recoil energy) and $\lambda$ (lattice wavelength) respectively. The Wannier functions are then evaluated with a harmonic expansion, allowing for the overlap of orbitals to change with variable $V_0$.

Taking $V_0=10E_R$, we find the orders of magnitude for the dipolar terms as $V = 10^{-3} \gamma E_R$, $T = 10^{-4} \gamma E_R$, and $P = 10^{-7} \gamma E_R$.
To gain better control over the long-range terms in an experimental scenario, a suitable combination of interaction processes would be required, i.e multiple attractive and repulsive long-range interactions or a large enough $g$ such that offsite contact terms are possible \cite{Dutta2015,Luhmann_2012}. Alternatively a setup exploiting light-matter processes to induce synthetic interactions \cite{Caballero_Benitez_2016,PhysRevLett.115.243604,Dogra2016,Niederle2016,landig2016,Flottat2017} could potentially be more efficient. However, in order to understand the effects of each process individually in this work, we will consider the parameters of Hamiltonian~(\ref{eq_ebh}) to be independent variables. This serves to show the effect of each term on the ground state.

\section{Self-consistent Gutzwiller mean-field} \label{_sc3}

In order to study the ground state phases that are possible within the general Bose-Hubbard model, we will consider a Gutzwiller mean-field approach. For interacting problems, the task of exact diagonalisation becomes unfeasible due to the exponential increase in the dimension of the Hilbert space. The Gutzwiller approach was first developed in the 60's for fermions \cite{Gutzwiller1963,Gutzwiller1964,Gutzwiller1965} and relies on approximating the many-body wave function by a product of on-site only contributions. This mean-field method was later extended to the case of bosons \cite{Rokhsar1991,Krauth1992} and applied to the Bose-Hubbard model of an ultracold bose gas in an optical lattice \cite{Jaksch1998}. The Gutzwiller approach relies on the assumption that quantum correlations are small \cite{lewenstein2012,pethick2008}, which is valid in the limit of large particle numbers and/or weak correlations between lattice sites (i.e. weak tunnelling). In the case of infinite dimensions, this treatment becomes exact; however, for lower dimensional cases, and especially for one dimension, there can be significant quantum correlations. While we will consider a one-dimensional model, we will work with a large particle number per site and small tunnelling strengths. In this limit, correlations between sites should be small and the Gutzwiller approach valid. It is also known that the Gutzwiller approach provides qualitatively correct results in one dimension but can not be trusted for the quantitative prediction of critical points of phase transitions, often over-estimating numerical values due to the neglection of correlations \cite{lewenstein2012,Zwerger2003}.

For bosonic atoms in an optical lattice, this mean-field approach is a suitable approximation to qualitatively capture the allowed phases since the majority of atoms will be in the condensed state. As such, operators may then be expressed as an average around some fluctuating operator,
\begin{equation}
\hat{b}_{i} \rightarrow \langle \hat{b}_{i} \rangle + \delta \hat{b}_{i},
\label{eq:averageOp}
\end{equation}
where $\delta \hat{b}_{i}$ denotes small deviations. Using Eq.~\eqref{eq:averageOp}, the nearest-neighbour summation terms in Hamiltonian~\eqref{eq_ebh} may then be decoupled to a problem which is on-site by linearising the fluctuation field. A natural extension of this mean-field is then to generalise the structure of the many-body wave function under these assumptions \cite{PhysRevA.71.043601}, which can be performed by using the Gutzwiller wave function,
\begin{equation} \label{eq_gpw}
| \Psi \rangle = \prod\limits_{i}^{L} \sum\limits_{n=0}^z f_n^{(i)} \ket{n_i},
\end{equation}
where $z$ is the maximum number of atoms allowed in each lattice site, $L$ denotes the size of the lattice, $\ket{n_i}$ is the state of $n$ atoms in site $i$, and $f_n^{(i)} $ are the coefficients which denote the mean-field wave function (commonly referred to as the Gutzwiller coefficients) and they are normalised such that
\begin{equation}
\sum\limits_{n=0}^z | f_n^{(i)} |^2 = 1.
\end{equation}

The many-body wave function \eqref{eq_gpw} is a product of on-site states such that we can rewrite the operators in Hamiltonian~\eqref{eq_ebh} which are over the nearest-neighbours in terms of on-site only operators. For the single particle tunnelling this is given by
\begin{equation}
\hat{b}_{i}^\dagger \hat{b}_{j} = \langle \hat{b}_{i}^\dagger \rangle \hat{b}_{j} + \langle \hat{b}_{j} \rangle \hat{b}_{i}^\dagger - \langle \hat{b}_{i}^\dagger \rangle \langle \hat{b}_{j} \rangle,
\end{equation}
the nearest-neighbour interaction by
\begin{equation}
\hat{n}_{i} \hat{n}_{j} = \langle \hat{n}_{i} \rangle \hat{n}_{j} + \langle \hat{n}_{j} \rangle \hat{n}_{i} - \langle \hat{n}_{i} \rangle \langle \hat{n}_{j} \rangle,
\end{equation}
the pair tunnelling by
\begin{equation}
\hat{b}_{i}^{\dagger 2} \hat{b}_{j}^2 = \langle \hat{b}_{i}^{\dagger 2}  \rangle \hat{b}_{j} + \langle \hat{b}_{j}^2 \rangle \hat{b}_{i}^{\dagger 2}  - \langle \hat{b}_{i}^{\dagger 2}  \rangle \langle \hat{b}_{j}^2 \rangle,
\end{equation}
and the density-dependent tunnelling by
\begin{equation}
\begin{aligned}
\hat{b}_{i}^\dagger (\hat{n}_i + \hat{n}_j ) & \hat{b}_{j}  =  \langle \hat{b}_{i}^\dagger \hat{n}_i \rangle \hat{b}_{j} + \langle \hat{b}_{j} \rangle \hat{b}_{i}^\dagger \hat{n}_i +\, \langle \hat{n}_j \hat{b}_{j} \rangle \hat{b}_{i}^\dagger \\ & + \langle \hat{b}_{i}^\dagger \rangle \hat{n}_j \hat{b}_{j} - \langle \hat{b}_{i}^\dagger \hat{n}_i \rangle \langle \hat{b}_{j} \rangle - \langle \hat{n}_j \hat{b}_{j} \rangle \langle \hat{b}_{i}^\dagger \rangle.
\end{aligned}
\end{equation}
From taking these nearest-neighbour terms to on-site only terms, we can see that there are four distinct, independent expectation values that are required, $\langle \hat{b}_{i} \rangle$, $\langle \hat{n}_{i} \rangle$, $\langle \hat{b}_{i}^2 \rangle$, and $\langle \hat{n}_{i} \hat{b}_{i} \rangle$. It comes as no surprise that these four expectations denote the four order parameters of Hamiltonian~\eqref{eq_ebh}. The first order parameter characterises the single-atom transport properties which we will label as $\varphi_i$ and is given by
\begin{equation} \label{eq_ord_p1}
\varphi_i = \langle \hat{b}_{i} \rangle = \sum\limits_{n=0}^z \sqrt{n}f_{n}^{(i)} f_{n-1}^{*(i)} = \langle \hat{b}_{i}^\dagger \rangle^*.
\end{equation}
Next, there is an order parameter which denotes the density behaviour of a phase, which we will label as $ \rho_i$ and is given by
\begin{equation} \label{eq_ord_p2}
\rho_i = \langle \hat{n}_{i} \rangle = \sum\limits_{n=0}^z n | f_{n}^{(i)} |^2.
\end{equation}
There is also an order parameter which defines the pair transport properties of a phase, which we will label as $\chi_i$ and is denoted by
\begin{equation} \label{eq_ord_p3}
\chi_i = \langle \hat{b}_{i}^2 \rangle = \sum\limits_{n=0}^z \sqrt{n(n-1)}f_{n}^{(i)} f_{n-2}^{*(i)} = \langle \hat{b}_{i}^{\dagger 2} \rangle^* .
\end{equation}
Finally, there is an order-parameter which defines the density-dependent transport properties, which we will label as $\eta_i$ and is given by
\begin{equation} \label{eq_ord_p4}
\eta_i = \langle \hat{n}_{i} \hat{b}_{i} \rangle = \sum\limits_{n=0}^z \sqrt{n}(n-1)f_{n}^{(i)} f_{n-1}^{*(i)} = \langle \hat{b}_{i}^\dagger \hat{n}_{i} \rangle^*.
\end{equation}
Physically, the order parameters represent observables that can be measured in the system. Note, each order parameter is defined for each  local lattice site, giving a total of $L$ values for each individual order parameter. However, due to the homogeneous nature of the lattice we consider, each of the $L$ order parameters will usually be the same, or 2-periodic. 

We note that the transport order parameters are probabilistic measures of how quickly atoms are being transferred to different lattice sites. This can be interpreted as a dynamical wave function from a macroscopic point of view, despite the observables being static. For example, insulating states have no transport but have clearly defined density fillings for each site. The superfluid, on the other hand, has homogeneous density and transport across the lattice, which can be viewed as a flowing macroscopic wave function with zero viscosity.

With the given relations for the order parameters, Hamiltonian~\eqref{eq_ebh} can then be written as a sum of on-site, mean-field Hamiltonians,
\begin{equation}
\mathcal{\hat{H}} = \sum\limits_{i}^{L} \hat{H}_i
\end{equation}
with each on-site Hamiltonian being given by
\begin{widetext}
\begin{equation} \label{eq_locH}
\begin{aligned}
\hat{H}_i = & -J (\hat{b}_{i} \bar{\varphi}_i^* + \, \hat{b}_{i}^\dagger \bar{\varphi}_i - \varphi_i^* \bar{\varphi}_i) + \dfrac{U}{2} \hat{n}_{i} (\hat{n}_{i} - 1) - \mu \hat{n}_{i} + V \bar{\rho}_i (\hat{n}_{i} - \dfrac{\rho_i}{2}) \\ & + T(\hat{b}_{i}^\dagger \hat{n}_{i} \bar{\varphi}_i + \hat{n}_{i} \hat{b}_{i} \bar{\varphi}_i^* + \hat{b}_{i}^\dagger \bar{\eta}_i + \hat{b}_{i} \bar{\eta}_i^* -\, \bar{\eta}_i \varphi_i^* - \bar{\varphi}_i \eta_i^* )+ P (\hat{b}_{i}^2 \bar{\chi}_i^* + \hat{b}_{i}^{\dagger 2} \bar{\chi}_i - \chi_i^* \bar{\chi}_i),
\end{aligned}
\end{equation}
\end{widetext}
where we have denoted nearest-neighbour summations by $\bar{x}_i$, i.e.
\begin{equation}
\bar{x}_i = \sum\limits_{\langle i,j\rangle}x_j.
\end{equation}
In this form, the ground state phase diagrams can be determined using a variety of methods. We will consider the approach of a self-consistent loop. However, it is also possible to use an imaginary time propagation approach, and we have confirmed that our results are consistent with this.

\begin{figure}[b]
	\centering,
	\makebox[0pt]{\includegraphics[width=0.95\linewidth]{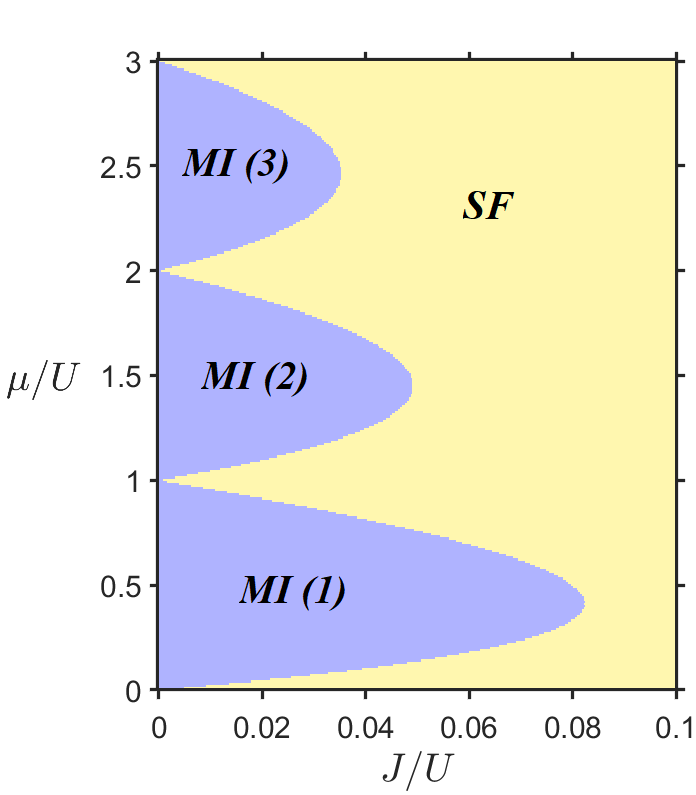}}
	\caption{Ground state phase diagram for the standard Bose-Hubbard model ($V/U = T/U = P/U = 0$), showing the Mott-insulator to superfluid phase transition.}
	\label{figure_x1}
\end{figure}

\begin{table*}
	\caption{List of phases, abbreviations and order parameter distributions.}
	\label{table_pm1}
	\centering
	\begin{tabular}{l c c c c c} \\
		\hline\hline \\,
		
		Phase & Abbreviation & $ \vec{\boldsymbol{\varphi}} $ & $ \vec{\boldsymbol{\rho}} $ & $ \vec{\boldsymbol{\eta}} $ & $ \vec{\boldsymbol{\chi}} $ \\ \\ [0.2ex]
		
		\hline \\
		
		\textit{Mott Insulator} & \textit{MI} & $ (0,0) $ & $ (\rho,\rho) $ & $ (0,0) $ & $ (0,0) $ \\
		\textit{Superfluid} & \textit{SF} & $ (\varphi,\varphi) $ & $ (\rho,\rho) $ & $ (\eta,\eta) $ & $ (\chi,\chi) $ \\
		\textit{Density Wave} & \textit{DW} & $ (0,0) $ & $ (\rho_a,\rho_b) $ & $ (0,0) $ & $ (0,0) $ \\
		\textit{Supersolid} & \textit{SS} & $ (\varphi_a,\varphi_b) $ & $ (\rho_a,\rho_b) $ & $ (\eta_a,\eta_b) $ & $ (\chi_a,\chi_b) $ \\
		\textit{One-body Staggered Superfluid} & \textit{OSSF} & $ (\varphi,-\varphi) $ & $ (\rho,\rho) $ & $ (\eta,-\eta) $ & $ (\chi,\chi) $ \\
		\textit{One-body Staggered Supersolid} & \textit{OSSS} & $ (\varphi_a,-\varphi_b) $ & $ (\rho_a,\rho_b) $ & $ (\eta_a,-\eta_b) $ & $ (\chi_a,\chi_b) $ \\
		\textit{Pair Superfluid} & \textit{PSF} & $ (0,0) $ & $ (\rho,\rho) $ & $ (0,0) $ & $ (\chi,-\chi) $ \\
		\textit{Pair Supersolid} & \textit{PSS} & $ (0,0) $ & $ (\rho_a,\rho_b) $ & $ (0,0) $ & $ (\chi_a,-\chi_b) $ \\
		\textit{Staggered Phases} & \textit{SP} & $ (\varphi_a,\varphi_b) $ & $ (\rho_a,\rho_b) $ & $ (\eta_a,\eta_b) $ & $ (\chi_a,-\chi_b) $ \\
		\textit{Intermediate Staggered Superfluid} & \textit{ISSF} & $ (\varphi_a,\varphi_b) $ & $ (\rho,\rho) $ & $ (\eta_a,\eta_b) $ & $ (\chi_a,-\chi_b) $ \\
		\textit{Staggered Superfluid} & \textit{SSF} & $ (\varphi_a,\varphi_b) $ & $ (\rho,\rho) $ & $ (\eta_a,\eta_b) $ & $ (\chi,-\chi) $ \\
		\textit{Staggered Supersolid} & \textit{SSS} & $ (\varphi_a,\varphi_b) $ & $ (\rho_a,\rho_b) $ & $ (\eta_a,\eta_b) $ & $ (\chi_a,-\chi_b) $ \\ [1.5ex]
		
		\hline

	\end{tabular}
	
\end{table*}

The self-consistent loop uses an exact diagonalisation scheme for the on-site problem. In other words, this amounts to solving a set of single-site problems coupled to one another through a mean-field. We initialise the loop by taking random and uniformly distributed order parameters for each site in the range $[0 , 1]$. Then, the local Hamiltonians are diagonalised such that the order parameters can be redefined from the ground state. Therefore, each local order parameter is updated when each local Hamiltonian is solved. After diagonalising the $L$ local Hamiltonians and updating the corresponding order parameters, the loop is then repeated until the energy and order parameters have converged to a given accuracy. We will converge to an accuracy of $10^{-4}$ in this study. From our results, the mean-field approach is stable, with it being rare that the convergence gets stuck in local minima corresponding to excited states. All phases discussed in this work have been checked for a number of iterations of the random initial order parameters to ensure that the phase diagrams are reflecting the ground state properties of the system.

\section{Characterisation of phases}
\label{sec:Character}

Before discussing the full phase diagrams, we will first outline the individual phases that will appear and their relation to the four order parameters defined in Sec.~\ref{_sc3}. We summarise all phases that are discussed in this work in Table.~\ref{table_pm1}, with vectors defining the corresponding order parameters for the lattice, i.e $ \vec{\boldsymbol{\varphi}} = (\varphi_1, \varphi_2, \varphi_3, \varphi_4, ... \, \varphi_L) $, where $\varphi_i$ is the order parameter for site $i$. For the considered phases, there are at most two unique terms for each set of order parameters, for example: $ \vec{\boldsymbol{\varphi}} = (\varphi_a, \varphi_b, \varphi_a, \varphi_b, ... \, \varphi_a, \varphi_b) $. Instead of writing the full vector $\vec{\boldsymbol{\varphi}}$ for all $L$ sites, we instead define $\vec{\boldsymbol{\varphi}}$ in an effective, compact periodic form as $ (\varphi_a, \varphi_b, \varphi_a, \varphi_b, ... \, \varphi_a, \varphi_b) \equiv (\varphi_a, \varphi_b)$. Note, we are not solving the effective 2 site problem, but are just using an alternative notation for the full lattice. In the standard Bose-Hubbard model, phases are defined by homogeneous order parameters. The translational symmetry in Hamiltonian~\eqref{eq_ebh} is broken by the two-site two-body terms, allowing for phases with two-site periodic order parameters.

For certain parameter regions, it is expected that we will observe the known Mott-insulator and superfluid phases. The Mott-insulator (MI) is defined by its fixed dynamics (no transport) and an integer valued uniform density across the lattice, i.e. $\vec{\boldsymbol{\varphi}} = \vec{\boldsymbol{0}}$ and $\rho_i \in \mathbb{Z}$ \cite{Fisher1989}. However, the superfluid (SF) phase is given by its uniform non-zero transport property and uniform non-integer density, i.e. $\vec{\boldsymbol{\varphi}} \neq \vec{\boldsymbol{0}}$ and $\vec{\boldsymbol{\rho}} \neq \vec{\boldsymbol{0}}$. It is known that the introduction of nearest-neighbour interactions can result in density wave (DW) and supersolid (SS) phases \cite{Rossini_2012,PhysRevB.95.195101,PhysRevB.86.054520}. The density wave phase is characterised by zero transport properties and a staggered (2-period) density, i.e. $\vec{\boldsymbol{\varphi}} = \vec{\boldsymbol{0}}$ and $\vec{\boldsymbol{\rho}} = \left(\rho_a,\rho_b\right)$, whereas the supersolid phase is defined by both staggered transport properties and staggered density, i.e. $\vec{\boldsymbol{\varphi}} = \left(\varphi_a,\varphi_b\right)$ and $\vec{\boldsymbol{\rho}} = \left(\rho_a,\rho_b\right)$.

Of course, the presence of the density-dependent and two-body pair tunnelling terms introduce more exotic phases. These phases are a result of the translational symmetry breaking two-site terms, including the density-dependent tunnelling, and the introduction of two-body dynamics by the pair tunnelling. Such transport staggered and pair superfluid phases have been previously observed for Hamiltonian~\eqref{eq_ebh} \cite{Maik_2013}. The one-body staggered superfluid (OSSF) is characterised by a sign staggering in the one-body transport properties, a constant two-body transport, and a constant non-integer density, i.e. $\vec{\boldsymbol{\varphi}} = \left( \varphi,-\varphi \right)$, $\vec{\boldsymbol{\rho}} \neq \vec{\boldsymbol{0}}$, $\vec{\boldsymbol{\eta}} = \left( \eta,-\eta\right)$ and $\vec{\boldsymbol{\chi}} \neq \vec{\boldsymbol{0}}$ . Similar to the ordinary supersolid, the one-body staggered supersolid (OSSS) has staggered one- and two-body transport, and a staggered density, however, the staggering is no longer only characterised by a sign flip. Additionally, it is possible to observe the pair superfluid and supersolid (PSF and PSS), which have no single-particle transport properties and a non-zero two-body transport. The dynamics of these phases are therefore dominated by two-body processes.

In this paper, we observe a new unconventional set of phases, which we will label as staggered phases (SP). These phases are the most general of phase, with all four order parameters staggered. The region of staggered phases is mainly made up of two separate phases with different symmetries in the density which exhibit an extended second-order (continuous) phase transition between them \cite{Landau1958}. The staggered supersolid (SSS) phase has all four order parameters staggered, whereas the superfluid phase (SSF) has a constant density.  Note, that as the transition between these two phases is of second order across a large region in parameter space, the labelling of a staggered superfluid and staggered supersolid phase is not clear in intermediate regions, and we can not rule out the possibility of multiple other phases existing. These regions of neither staggered superfluid or staggered supersolid phase could be a result of the second-order phase transition or the signature of more exotic phases. We expect that these staggered phases exhibited by the Gutzwiller mean-field could be related to the previously observed twisted complex staggered phases in DMRG studies \cite{srep12912,Luhmann2016}. However, an extension of the mean-field approach used here would be required to compare the staggered and twisted phases, as the twisted phases exhibit complex order parameters, which is outside the scope of this work. This is due to the nature of the static Gutzwiller mean-field approach, where non-local correlations are neglected. Therefore, care is required when considering initial states with phase differences between local lattice sites.

\begin{figure*}
\centering
\makebox[0pt]{\includegraphics[width=0.95\textwidth]{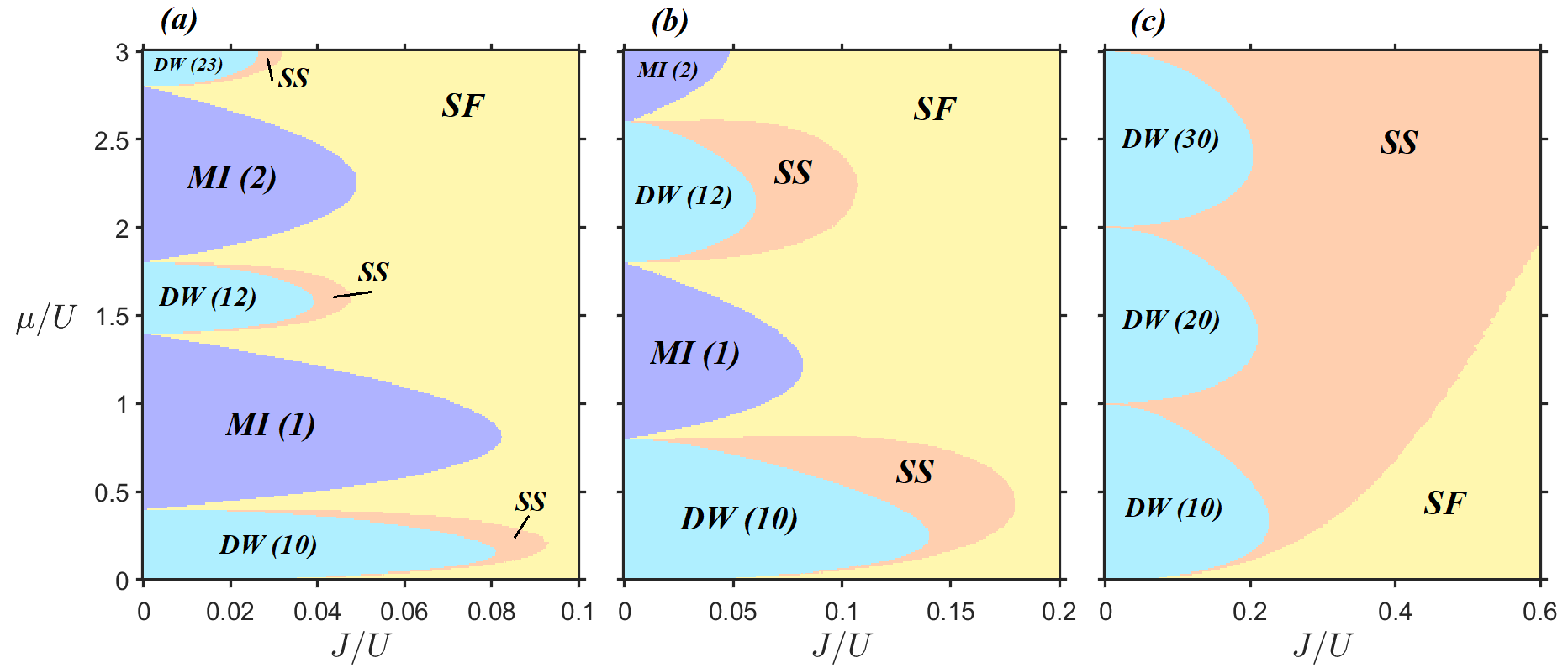}}
\caption{Ground state phase diagrams with nearest-neighbour interactions present in the system, revealing the presence of inversion symmetry breaking DW and SS phases. We consider the cases of (a) weak neighbour interactions ($V/U = 0.2, T/U = P/U = 0$), (b) intermediate neighbour interactions ($V/U = 0.4, T/U = P/U = 0$) and (c) strong neighbour interactions ($V/U = 0.8, T/U = P/U = 0$). The bracketed indices next to the DW abbreviation corresponds to the expectation value of the number operators ($\rho_a$, $\rho_b$) across two sites.}
\label{figure_x2}
\end{figure*}

\begin{figure*}
	\centering
	\makebox[0pt]{\includegraphics[width=0.95\textwidth]{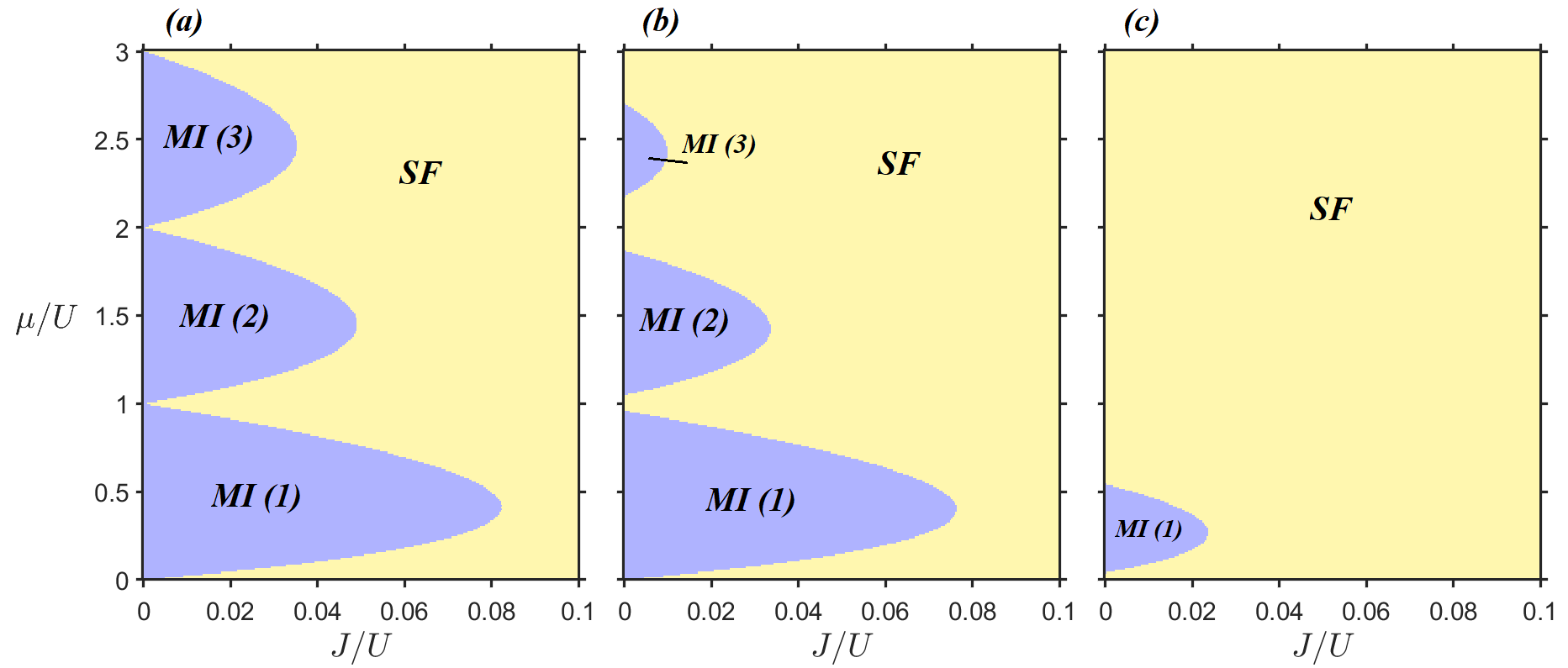}}
	\caption{Ground state phase diagrams with density-dependent tunnelling present in the system, considering the cases of (a) local interactions ($V/U = T/U = P/U = 0$), (b) weak density-dependent tunnelling ($V/U = P/U = 0, \, T/U = -0.005$) and (c) strong density-dependent tunnelling ($V/U = P/U = 0, \, T/U = -0.05$). The Mott-insulating phases are observed to be destroyed for increasing $T/U$. }
	\label{figure_x3}
\end{figure*}

\begin{figure*}
	\centering
	\makebox[0pt]{\includegraphics[width=0.95\textwidth]{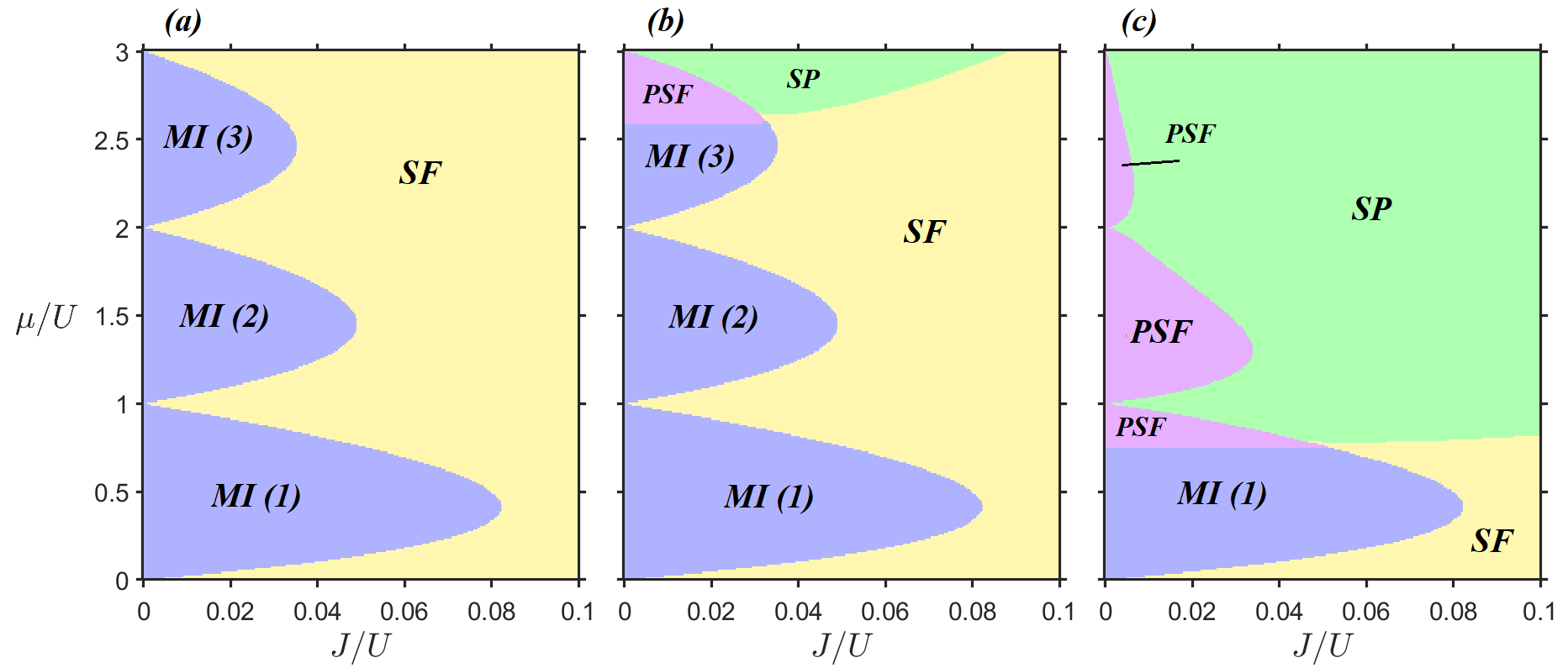}}
	\caption{Ground state phase diagrams with pair tunnelling present in the system, considering the cases of (a) local interactions ($V/U = T/U = P/U = 0$), (b) weak pair tunnelling ($V/U = T/U = 0, \, P/U = 0.035$) and (c) strong pair tunnelling ($V/U = T/U = 0, \, P/U = 0.12$). In a similar manner to the density-dependent tunnelling case, the extent of insulating domains is again reduced even for modest $P/U$.}
	\label{figure_x4}
\end{figure*}

\begin{figure*}[!]
	\centering
	\makebox[0pt]{\includegraphics[width=0.95\textwidth]{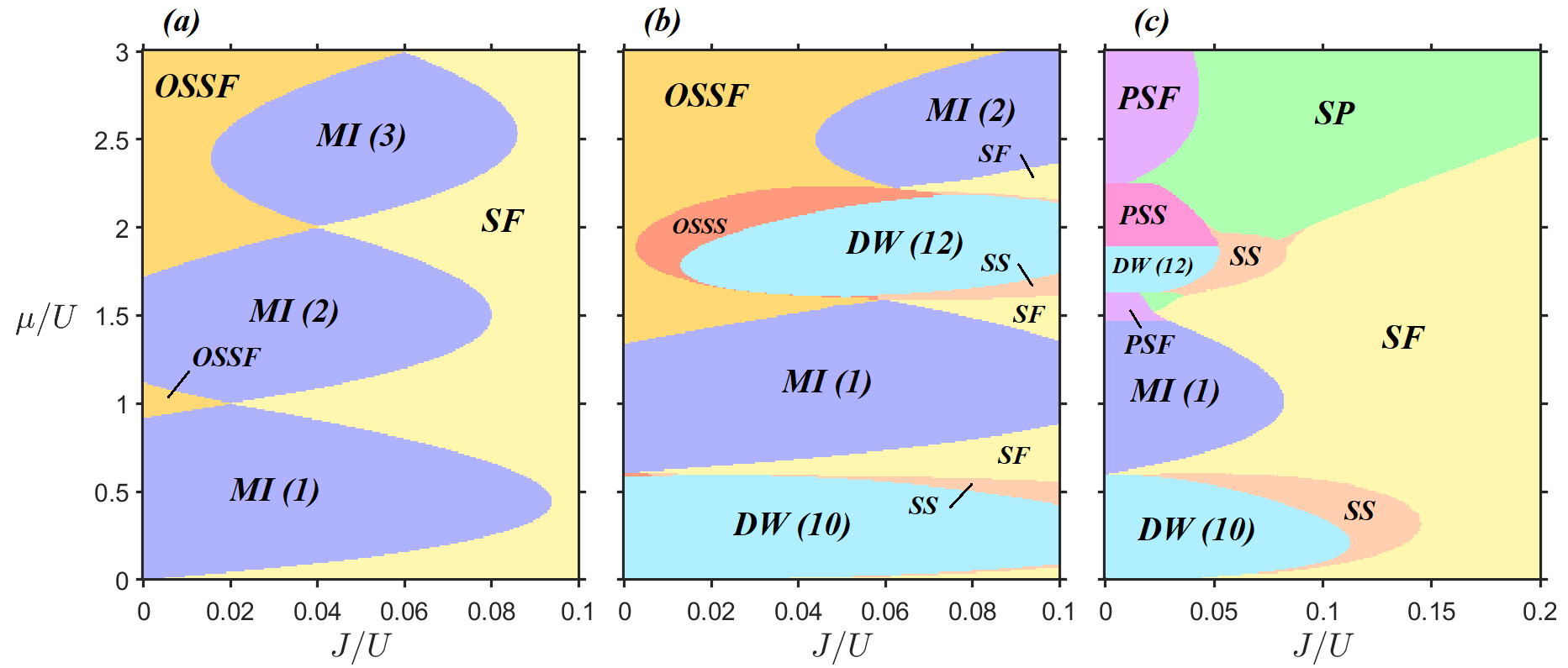}}
	\caption{Ground state phase diagrams with combinations of various interaction processes of interest. We consider the cases of (a) opposite sign density-dependent tunnelling ($V/U = P/U = 0, \, T/U = 0.01$), (b) opposite sign density-dependent tunnelling with neighbour interactions ($V/U = 0.3, \, T/U = 0.03, \, P/U = 0$) and (c) pair tunnelling with neighbour interactions ($V/U = 0.3, \, T/U = 0, \, P/U = 0.1$). The emergence of novel superfluid phases such as the OSSF, SP and PSF can be seen within certain regions. Furthermore, when nearest-neighbour interactions are present, additional long-range, supersolid phases are found (OSSS and PSS).}
	\label{figure_x5}
\end{figure*}

\section{Ground State Phase Diagrams} \label{_sc4}

In this section, mean-field phase diagrams are presented and discussed for various parameter regimes of Hamiltonian~\eqref{eq_ebh}. To accurately resolve the phase boundaries, we use a grid of at least $2500$ points for each phase diagram. We will work in units of the two-body on-site interaction strength $U$ and consider one-dimensional equally spaced lattices of $10$ sites with periodic boundary conditions. The size of the truncated basis, denoted by the maximum number of particles per site $z$, is selected so that the convergence of order parameters is constant with respect to the desired convergence precision when $z$ is increased. We have found that in the considered $10$ site lattice, a maximum particle number of $z=20$ is required to have machine precision in convergence.

First, we check that for the case of $V/U = T/U = P/U = 0$, the well-known Mott-insulator to superfluid transition is observed. We indeed see this behaviour in Fig.~\ref{figure_x1}, with the distinct Mott-insulator lobes which are characterised by different integer uniform fillings of the lattice sites. The critical point of the Mott-insulator to superfluid transition is found to be in agreement with previous Gutzwiller mean-field approaches \cite{Zwerger2003,lewenstein2012}, with a critical transition point of $(J/U)_c = 0.0825$ for the first order Mott-insulator. As expected from mean-field results this is an over-estimation of the true critical point \cite{Kashurnikov1996,Kuhner2000}.

We also confirm that given a non-zero $V/U$, we observe the known supersolid and density wave phases, as shown in Fig.~\ref{figure_x2}. For sufficiently strong $V/U$, the Mott-insulating phases are completely destroyed and replaced with the density wave phase. This transition to density wave phases makes sense, as the nearest-neighbour interactions can be reduced by generating an offset between the density of nearest-neighbours. Therefore, there is a symmetry breaking of the ground state in order to reduce its energy. It is also observed that the higher order density wave phases exist over a consistent size of parameters, i.e. the area of the density phase is similar for different orders of the density wave. With increasing $V/U$ the superfluid phase is also replaced but with a supersolid. The supersolid phase exists because of the same symmetry arguments already invoked for the density wave, but starting from a superfluid. 

\begin{figure*}[t]
	\centering
	\makebox[0pt]{\includegraphics[width=0.95\textwidth]{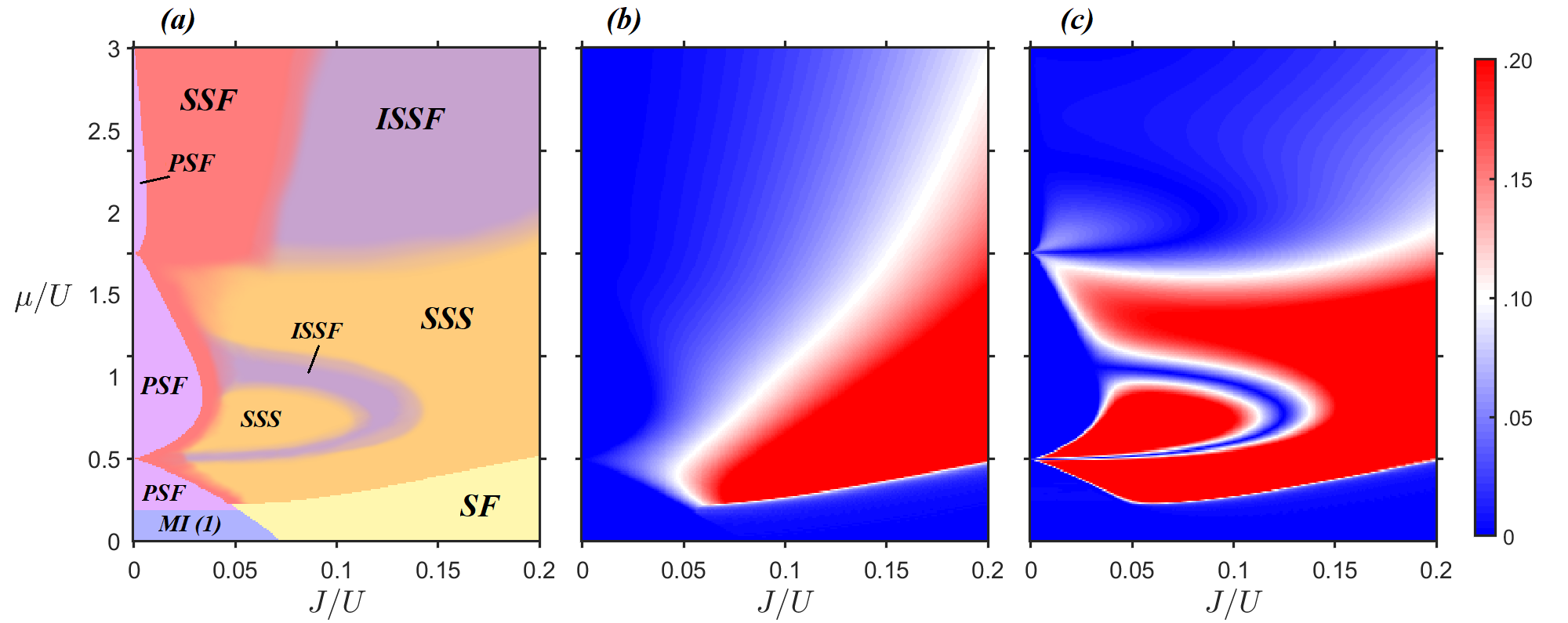}}
	\caption{Phase diagrams and modulation order parameters when $V/U = T/U = 0$ and $P/U = 0.12$. (a) The detailed staggered phase region of Fig.~\ref{figure_x4}, showing the dominant phases with second-order phase transitions between them. We plot the (b) pair ($||\chi_1|-|\chi_2||$) and (c) density ($|\rho_1-\rho_2|$) modulations, which characterise the three phases within the SP region. The colour maps of the pair and density modulation are normalised to $1$ from maximum values of $0.7154$ and $0.0411$ respectively. To highlight the structure of the phases, the colour maps are saturated to $0.2$. }
	\label{figure_x6}
\end{figure*}

A non-zero density-dependent tunnelling causes the destruction of the Mott-insulator phase, as we observe in Fig.~\ref{figure_x3}. This is due to there being an incentive for the state to spread out its density, and favours the superfluid. That is, it is energetically favourable for the density of the ground state to be lower than that required for the Mott-insulator state. With $T/U \neq 0$, the magnitude of density-dependent tunnelling enlarges, thus reducing the stability of insulating phases and inflating the overall tunnelling rate. 

If there is a non-zero pair tunnelling strength, then the non-trivial pair superfluid and staggered phases are observed, as seen in Fig.~\ref{figure_x4}. The pair superfluid effectively replaces the Mott-insulating lobes at high enough chemical potential. This makes sense, as the system now has a non-zero pair tunnelling, and hence, the pair tunnelling order parameter can not remain zero at large chemical potential. The staggered phases arise mostly in the region of the phase diagram that usually consists of a superfluid. Due to the process of pair tunnelling, it is more likely for large $P/U$ that density and transport properties will clump together and favour some lattice sites over the others. This asymmetry in the density and transport properties results in the staggered phases, where all order parameters are in general staggered.

In Fig.~\ref{figure_x5}a, we study a modification of the density dependent tunnelling process with the opposite sign to the linear tunnelling strength, corresponding to the presence of an attractive long-range interaction. This leads to a skewed structure of the phase diagram, with the Mott-insulating lobes looking more like elliptical structures for large $\mu/U$. Furthermore, both a superfluid and one-body staggered superfluid phase are shown to exist. The one-body staggered superfluid phase arises in regions where $T/U \sim - J/U$, i.e. where the attractive non-local interactions are dominating the process, and hence a breaking of the translational symmetry is energetically favourable.

Combinations of several processes are then tuned within Figs.~\ref{figure_x5}b and~\ref{figure_x5}c in order to display the long-range (crystalline order) counterparts of the unconventional one-body staggered superfluid and pair superfluid phases. In Fig.~\ref{figure_x5}b, both neighbour interactions and opposite sign density-dependent tunnelling are considered simultaneously. This leads to the familiar inclusion of supersolid and density wave phases that were already considered. However, we also observe a small parameter region where the one-body staggered supersolid phase exists. This phase is due to the staggering of the transport order parameters being favourable but the tunnelling being strong enough and the chemical potential small enough to not yet favour the one-body staggered superfluid phase. To access this particular phase diagram in a physical set-up, one would require the presence of both repulsive and attractive interactions simultaneously (with suitable tuning). Finally, in Fig.~\ref{figure_x5}c we again demonstrate density wave and supersolid phases for when both neighbour interactions and pair tunnelling are present. We also observe the non-trivial phases of the pair superfluid and supersolid due to the favouring of pair transport for large numbers of atoms (large $\mu/U$), with no single-particle transport present in the ground states. 

From the considered phase diagrams, we know that negligible $T/U$ is required in order to stabilise the long-range, symmetry breaking staggered phases. In particular, we must have $|P|/U > |T|/U$ to observe the staggered phases which, in a purely dipolar setup, requires carefully balanced repulsive and attractive long-range interactions as the relation given in Eq.~\eqref{eq:Scaling} is not satisfied. However, synthetic many-body processes induced by light-matter interactions \cite{Caballero_Benitez_2016,PhysRevLett.115.243604,Dogra2016,Niederle2016,landig2016,Flottat2017} could alternatively be used to control the allowed phase transitions with greater freedom in an experimental scenario.

It is also worthwhile to consider the staggered phase regions in more detail. We will focus here on the staggered phases appearing in Fig.~\ref{figure_x4}c, which has strong pair tunnelling and no long-range interactions or density-dependent tunnelling. In Fig.~\ref{figure_x6}, we consider the staggered phase and label regions of staggered supersolid, staggered superfluid, and an intermediate staggered superfluid (ISSF). All transitions between each staggered phase is second-order and in Fig.~\ref{figure_x6}a, we label regions where certain phases are dominant. The intermediate staggered superfluid is characterised by a constant density but with a pair transport which is staggered between different values, i.e. $\vec{\boldsymbol{\chi}} = \left( \chi_a,-\chi_b,\dots \right)$. This peculiar property of the intermediate phases are shown in the pair and density modulation order parameter plots of Fig.~\ref{figure_x6}b and Fig.~\ref{figure_x6}c respectively. The staggered superfluid is defined when both the pair and density modulation is equal to zero, whereas the staggered supersolid has finite modulation. In the intermediate case, there is finite pair modulation but zero density modulation.

It would be natural to think that the intermediate staggered superfluid is only a signature of the second-order phase transition. Regardless, it is interesting that in the second-order transition between the staggered superfluid and supersolid the density and pair order parameters appear to change their symmetry properties on different scales, resulting in the dominance of this third intermediate phase. However, it is observed that for moderately large $\mu/U$ and $J/U$, the dominant phase is the intermediate staggered superfluid. This phase is of a superfluid nature with a constant non-integer density and a non-zero staggered transport property. The symmetry breaking of the pair tunnelling is a result of the two-body nature of the pair tunnelling process. This phase is similar to the staggered superfluid but it is of a more general type, i.e. all staggered superfluids can also be classified as being in the intermediate phase but not all intermediate staggered superfluids are of the staggered superfluid type.

\section{Conclusions} \label{_sc5}

In this work, we have considered the ground state phases of long-range interacting bosonic ultracold gases in optical lattices. This model is the most general of Bose-Hubbard models for two-body interactions. In addition to the standard local terms, this model includes pair tunnelling, long-range two-body interactions, and a density-dependent (or induced) tunnelling. We consider the ground state phases of this Bose-Hubbard model via a Gutzwiller local mean-field approach, which is valid in the limit of small quantum correlations. Therefore, it is necessary to consider large Hilbert spaces and small tunnelling strengths to be in the region where the mean-field is valid. 

We confirm the presence of density wave and supersolid phases with intermediate and strong nearest-neighbour two-body interactions. In addition, we observe the known destruction of the Mott-insulating phase by a density-dependent tunnelling process. By considering the behaviour of the system with non-zero pair tunnelling, we observe an interesting mixed-phase region where all order parameters, or all but the density, are staggered. This region consists of a mixture between superfluid and supersolid phases, with second-order phase transitions between them over a large region of parameter space. 

In summary, we have provided a detailed study of the ground state phases of the general Bose-Hubbard model for long-range interacting atoms. We observed the existence of new,  2-site periodic phases in the 1D limit. From the corresponding order parameters, we have observed not only unconventional sign staggering for both one-body and pair superfluids, but a rich, non-trivial structure in the symmetry transitions of the staggered phases. While these results are general, and not specific to an atomic species or set-up, we expect that the non-trivial ground states observed here could play an important role for dipolar atomic gases in optical lattices. In particular, the new staggered phases could be observed when dipolar gases are combined with synthetic techniques, e.g. light-matter interactions \cite{Caballero_Benitez_2016,PhysRevLett.115.243604,Dogra2016,Niederle2016,landig2016,Flottat2017}, to induce competing long-range many-body interactions.

\begin{acknowledgments}
D.J., N.W. and C.W.D. acknowledge support from EPSRC CM-CDT Grant No. EP/L015110/1. P.\"{O}. acknowledges support from EPSRC Grant No. EP/M024636/1.
\end{acknowledgments}



%

\end{document}